\documentclass[preprint]{aastex63}

\newcommand{\fe}{\ion{Fe}{1}}
\newcommand{\fea}{\ion{Fe}{1} $\lambda$6569}
\newcommand{\ha}{H$\alpha$}
\newcommand{\haa}{H$\alpha$ $\lambda$6563}

\usepackage{color}
\usepackage{multirow}
\usepackage{amsmath}

\begin{document}

\title{The White-light Emissions in Two X-class Flares Observed by ASO-S and CHASE}

\author[0000-0002-8258-4892]{Ying Li}
\author[0000-0002-8401-9301]{Zhichen Jing}
\affiliation{Key Laboratory of Dark Matter and Space Astronomy, Purple Mountain Observatory, CAS, Nanjing 210023, People's Republic of China}
\affiliation{School of Astronomy and Space Science, University of Science and Technology of China, Hefei 230026, People's Republic of China}

\author[0000-0003-0057-6766]{De-Chao Song}
\affiliation{Key Laboratory of Dark Matter and Space Astronomy, Purple Mountain Observatory, CAS, Nanjing 210023, People's Republic of China}

\author[0000-0001-7540-9335]{Qiao Li}
\author[0000-0002-1068-4835]{Jun Tian}
\author[0000-0002-3657-3172]{Xiaofeng Liu}
\affiliation{Key Laboratory of Dark Matter and Space Astronomy, Purple Mountain Observatory, CAS, Nanjing 210023, People's Republic of China}
\affiliation{School of Astronomy and Space Science, University of Science and Technology of China, Hefei 230026, People's Republic of China}

\author[0000-0003-3699-4986]{Ya Wang}
\affiliation{Key Laboratory of Dark Matter and Space Astronomy, Purple Mountain Observatory, CAS, Nanjing 210023, People's Republic of China}

\author[0000-0002-4978-4972]{M. D. Ding}
\affiliation{School of Astronomy and Space Science, Nanjing University, Nanjing 210023, People's Republic of China}
\affiliation{Key Laboratory for Modern Astronomy and Astrophysics (Nanjing University), Ministry of Education, Nanjing 210023, People's Republic of China}

\author[0000-0003-4490-7344]{Andrea Francesco Battaglia}
\affiliation{University of Applied Sciences and Arts Northwestern Switzerland, Bahnhofstrasse 6, 5210 Windisch, Switzerland}
\affiliation{ETH Z{\"u}rich, R{\"a}mistrasse 101, 8092 Z{\"u}rich, Switzerland}

\author[0000-0003-4655-6939]{Li Feng}
\author[0000-0003-1078-3021]{Hui Li}
\affiliation{Key Laboratory of Dark Matter and Space Astronomy, Purple Mountain Observatory, CAS, Nanjing 210023, People's Republic of China}
\affiliation{School of Astronomy and Space Science, University of Science and Technology of China, Hefei 230026, People's Republic of China}

\author{Weiqun Gan}
\affiliation{Key Laboratory of Dark Matter and Space Astronomy, Purple Mountain Observatory, CAS, Nanjing 210023, People's Republic of China}
\affiliation{University of Chinese Academy of Sciences, Nanjing 211135, People's Republic of China}

\correspondingauthor{Ying Li}
\email{yingli@pmo.ac.cn}
\correspondingauthor{De-Chao Song}
\email{dcsong@pmo.ac.cn}

\begin{abstract}
The white-light continuum emissions in solar flares (i.e., white-light flares) are usually observed on the solar disk but, in a few cases, off the limb. Here we present on-disk as well as off-limb continuum emissions at 3600 \AA\  (in the Balmer continuum) in an X2.1 flare (SOL2023-03-03T17:52) and an X1.5 flare (SOL2023-08-07T20:46), respectively, observed by the White-light Solar Telescope (WST) on the Advanced Space-based Solar Observatory (ASO-S). These continuum emissions are seen at the ribbons for the X2.1 flare and on loops during the X1.5 event, in which the latter also appears in the decay phase. These emissions also show up in the pseudo-continuum images at \fe\ $\lambda$6173 from the Helioseismic and Magnetic Imager (HMI) on the Solar Dynamics Observatory (SDO). In addition, the ribbon sources in the X2.1 flare exhibit significant enhancements in the \fe\ line at 6569.2 \AA\ and the nearby continuum observed by the Chinese \ha\ Solar Explorer (CHASE). It is found that the on-disk continuum emissions in the X2.1 flare are related to a nonthermal electron-beam heating either directly or indirectly, while the off-limb emissions in the X1.5 flare are associated with thermal plasma cooling or due to Thomson scattering. These comprehensive continuum observations can provide good constraints on flare energy deposition models, which helps well understand the physical mechanism of white-light flares.
\end{abstract}



\section{Introduction} 
\label{sec:intro}

The first solar flare observed in history, i.e., the Carrington event \citep{Carrington1859,Hodgson1859} was a white-light flare (WLF) characterized by an enhancement in the visible continuum. It has been proposed that the white-light continuum enhancement is supposed to be present in all flares but depending on instrumental sensitivity \citep{Neidig1989,Hudson2006}. So far there are only several hundred solar WLFs reported in the literature based on ground- and space-based observations. Generally speaking, the larger flares, say, with X- and M-class, are more likely identified as WLFs \citep[e.g.,][]{Song2018,Jing2024}.

The emission sources and heating mechanisms of WLFs have been explored for a few decades. Negative hydrogen (H$^{-}$) emission from the photosphere and hydrogen recombination continuum from the chromosphere are considered to be the main sources of white-light emission \citep{Hiei1982,Ding2007}. In addition, Thomson scattering can dominantly contribute to the white-light emission in the corona when electron density is smaller than 10$^{12}$ cm$^{-3}$ \citep{Hiei1992,Fang1995,Heinzel2017}. Note that the hydrogen free-free emission can also contribute to some of the white-light continuum \citep{Heinzel2017}. For the heating mechanisms, there have been proposed four types: (1) beams of electron \citep{Hudson1972,Fletcher2007,Krucker2015} and proton \citep{Machado1978,Prochazka2018}, (2) irradiation from X-rays \citep[e.g.,][]{Henoux1977} or EUV \citep[e.g.,][]{Poland1988} and radiative backwarming from chromosphere \citep[e.g.,][]{Ding2003,Xu2004},  (3) chromospheric condensation \citep{Gan1992,Kowalski2015}, (4) Alfv\'{e}n waves \citep{Emslie1982,Fletcher2008}. Note that in a certain flare, multiple mechanisms could play roles in producing the white-light emission \citep[e.g.,][]{Machado1978,Xu2006,Hao2017,Song2023}.

Both on-disk and off-limb observations of the white-light emission are valuable for understanding the physical mechanism of WLFs. On-disk white-light emissions have been observed in majority of the WLFs reported before, while the off-limb continuum sources are rarely studied. At early times, WLFs were mainly observed in the Paschen continuum below 8203 \AA\ \citep[e.g.,][]{McIntosh1972,Hiei1992} from ground-based telescopes, along with some events observed in the Balmer continuum below 3646 \AA\ \citep[e.g.,][]{Hiei1982,Neidig1983}. In recent years, the Helioseismic and Magnetic Imager (HMI; \citealt{Scherrer2012}) on the Solar Dynamics Observatory (SDO; \citealt{Pesnell2012}) has detected more than one hundred WLFs via the \fe\ $\lambda$6173 line. In particular, some off-limb WLFs were reported and their mechanisms were investigated \citep{Oliveros2014,Heinzel2017,Fremstad2023,Zhao2023}. By using the NUV data from the Interface Region Imaging Spectrograph (IRIS; \citealt{Pontieu2014}), several on-disk WLFs were studied in detail in the Balmer continuum \citep{Heinzel2014,Kleint2016,Joshi2021}. However, the Balmer continuum from off-limb WLFs still lacks of evidence in observations.

The newly launched missions of the Advanced Space-based Solar Observatory (ASO-S; \citealt{Gan2023}) and the Chinese \ha\ Solar Explorer (CHASE; \citealt{LiC2022}) carry instruments that observe the Sun in the wave band at 3600 \AA\ (i.e., the Balmer continuum) and in the spectral lines of \fea\ and \haa, respectively. Here we present on-disk as well as off-limb continuum emissions at 3600 \AA\ and near the \fe\ $\lambda$6173/$\lambda$6569 line (in the Paschen continuum) in two X-class flares. These comprehensive white-light observations can provide good constraints on the energy transportation and deposition in WLFs.


\section{Observational Data}
\label{sec:data}

Multi-wavelength images and visible spectra are analyzed in this study. CHASE provides the two-dimensional spectra of \fe\ at 6569.2 \AA\ and \ha\ at 6562.8 \AA\ for the full Sun, whose spectral, spatial, and temporal resolutions are 0.05 \AA\ pixel$^{-1}$, $\sim$1\arcsec, and $\sim$1 min, respectively, with a two-binning mode in spectrum and space \citep{Qiu2022}. The White-light Solar Telescope (WST) on the Ly$\alpha$ Solar Telescope (LST; \citealt{LiH2019}) carried by ASO-S provides the white-light continuum images at 3600 \AA\ for the full disk. The images have a spatial resolution of $\sim$4\arcsec\ (though the pixel size being 0.5\arcsec) and a cadence of 2 min in a routine mode. In a burst mode for flare events, the cadence can be high as 1 s. SDO/HMI obtains the pseudo-continuum images at \fe\ $\lambda$6173 with a spatial resolution of 0.5\arcsec\ pixel$^{-1}$ and a cadence of 45 s. The Atmospheric Imaging Assembly (AIA; \citealt{Lemen2012}) on SDO obtains the UV (1600 and 1700 \AA) and EUV (94, 131, 171, 193, 211, 304, and 335 \AA) images with a pixel size of 0.6\arcsec\ pixel$^{-1}$ and a cadence of 24 s or 12 s. An EUV image at 174 \AA\ (L2 data\footnote{https://doi.org/10.24414/z818-4163}) from the Full Sun Imager (FSI) of the Extreme Ultraviolet Imager (EUI; \citealt{Rochus2020}) on Solar Orbiter (SolO; \citealt{Muller2020}) is also presented here, which has a pixel size of 4.4\arcsec.

Some soft X-ray (SXR) and hard X-ray (HXR) data are used in this work as well. The SXR 1--8 \AA\ data are provided by the X-Ray Sensor (XRS; \citealt{Hanser1996}) on Geostationary Operational Environmental Satellite (GOES). The Gamma-ray Burst Monitor (GBM; \citealt{Meegan2009}) on Fermi obtains the HXR spectra with an energy range from 8 keV to 40 MeV. The data from detector N5 are used in this work. The Spectrometer Telescope for Imaging X-rays (STIX; \citealt{Krucker2020}) on SolO provides the HXR spectra at 4--150 keV and the L1 pixel data are used here for the second flare. Note that SolO had a separation angle of 152.8$^{\circ}$ with Earth and a distance of 0.88 AU to the Sun at the time of observation. The times from SolO observations have been corrected to Earth time.


\section{Observations and Results}
\label{sec:result}

\subsection{The on-disk X2.1 Flare on 2023 March 3}

\subsubsection{Overview of the Flare Event}

\begin{figure}[htb]
	\centering
	\includegraphics[width=0.8\textwidth]{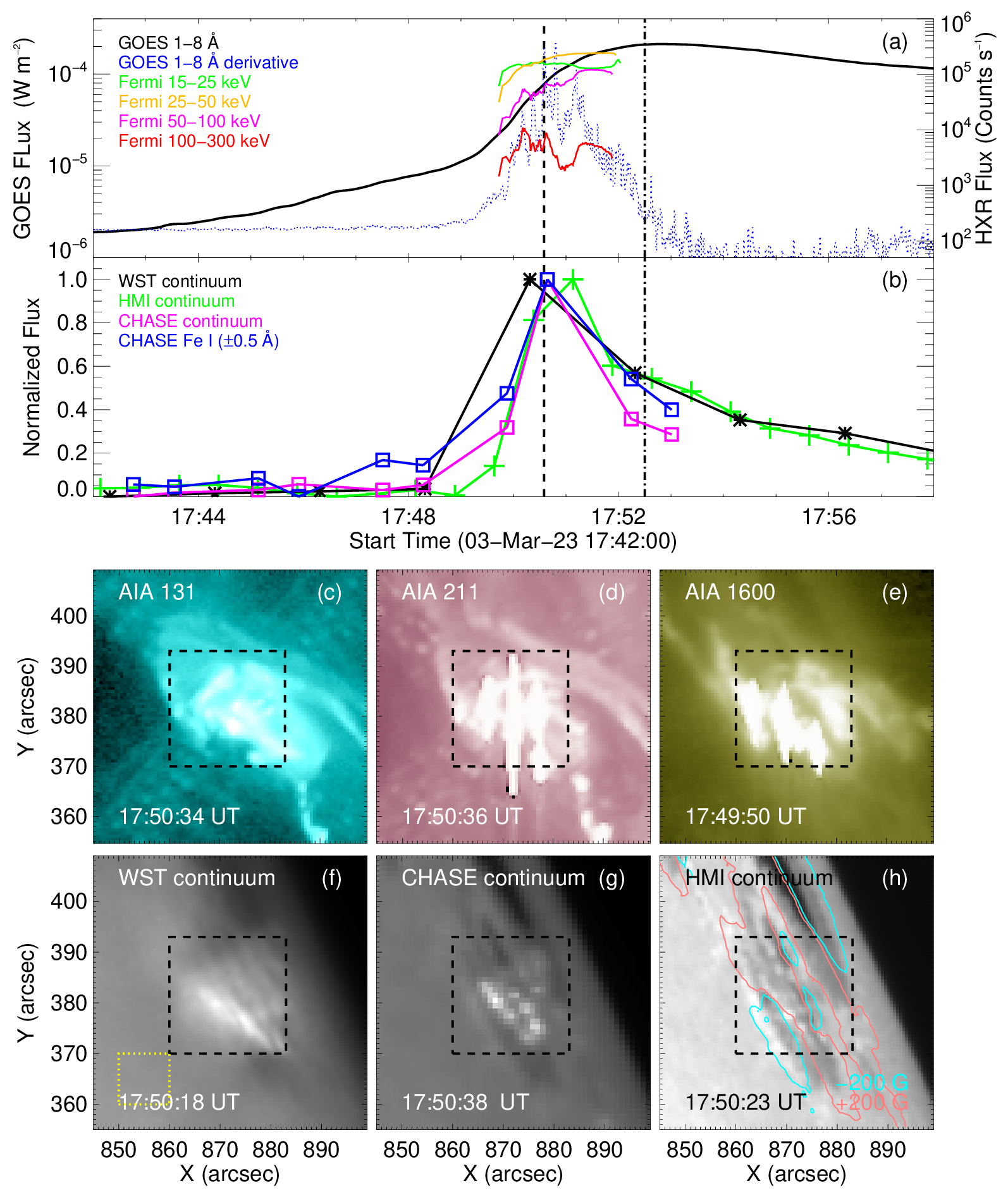}
	\caption{Overview of the on-disk X2.1 flare on 2023 March 3. (a) Light curves of GOES SXR 1--8 \AA\ flux, its time derivative, and available HXR fluxes from Fermi. (b) Emission curves of WST continuum at 3600 \AA, HMI continuum at $\sim$6173 \AA, CHASE continuum at $\sim$6569 \AA, and CHASE \fea\ line (integrated over $\pm$0.5 \AA) for the flaring region marked by the black dashed box in panels (c)--(h). The vertical dashed and dash-dotted lines in panels (a) and (b) indicate the peak times ($\sim$17:50 and $\sim$17:52 UT) of SXR time derivative and SXR 1--8 \AA\ flux, respectively. (c)--(h) Multi-wavelength images of the flaring AR at $\sim$17:50 UT. The black dashed box marks the flaring region that is integrated over for the emission curves in Figure \ref{fig1}(b). The yellow dotted box in panel (f) is used to estimate the uncertainties of white-light emission enhancements. The pink and cyan contours in panel (h) represent the positive and negative magnetic polarities at $\pm$200 G, respectively.}
	\label{fig1}
\end{figure}

The X2.1 flare (SOL2023-03-03T17:52) was located in the NOAA active region (AR) 13234 (N21W76), near the west limb but still on the disk. It started at 17:39 UT, peaked at 17:52 UT, and ended at 18:55 UT on 2023 March 3. The GOES SXR 1--8 \AA\ flux and its time derivative are plotted in Figure \ref{fig1}(a), together with the Fermi HXR fluxes from the flare. We can see that the HXR emission has a response up to 300 keV in the impulsive phase. In particular, the HXR emission at 100--300 keV seems to resemble the SXR time derivative well, indicative of the Neupert effect \citep{Neupert1968}. These indicate that a nonthermal electron-beam heating takes place during the flare. Figure \ref{fig1}(b) shows multiple continuum emission curves from WST at 3600 \AA, HMI at $\sim$6173 \AA, and CHASE at $\sim$6569 \AA, and also the \fea\ intensity curve (integrated over $\pm$0.5 \AA) for the flaring region. It is seen that all the light curves peak at $\sim$17:50 UT, i.e., around the same time with the peak of SXR time derivative, which also correspond to the HXR emission at 100--300 keV. Note that the cadences of WST and CHASE data are lower than 1 min and that CHASE observed this flare only till 17:53 UT. Figures \ref{fig1}(c)--(h) show multi-wavelength images for the flaring AR at $\sim$17:50 UT. One can see some loop structures in AIA 131 and 211 \AA\ images as well as bright ribbons (with different magnetic polarities) in the continuum images from WST, CHASE, and HMI. Note that the AIA 211 and 1600 \AA\ images are saturated significantly at the flare ribbons.

\subsubsection{The White-light Emissions at the Flare Ribbons}

\begin{figure}[htb]
	\centering
	\includegraphics[width=1.0\textwidth]{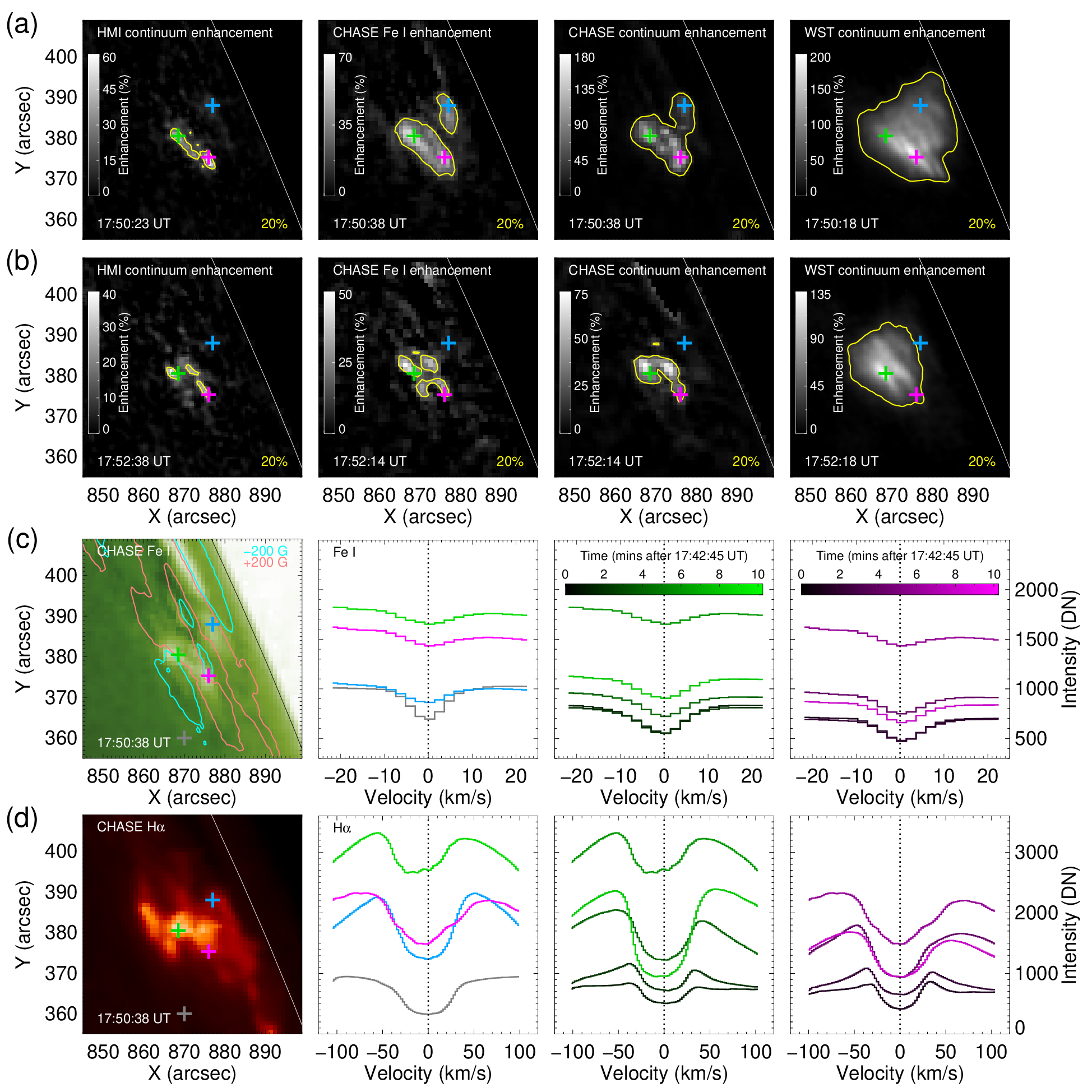}
	\caption{(a) and (b) Maps of relative enhancements of HMI continuum, CHASE \fe\ line center, CHASE continuum, and WST continuum emissions at $\sim$17:50 and $\sim$17:52 UT for the X2.1 flare. The solar limb is marked by a white line. In each panel, the yellow contours indicate an enhancement level of 20\% and the three colored plus symbols denote three representative brightening kernels. (c) and (d) Line profiles of CHASE \fe\ and \ha\ at three selected kernels at $\sim$17:50 UT (the second column, corresponding to the \fe\ and \ha\ intensity maps shown in the first column) as well as the temporal evolution of line profiles at the two most brightened kernels (the third and fourth columns, corresponding to the green and magenta pluses, respectively). A gray plus symbol in the first column indicates a quiet-Sun location where the \fe\ and \ha\ line profiles are displayed in the second column for comparison. The pink and cyan contours in the first column of panel (c) represent the positive and negative magnetic polarities at $\pm$200 G, respectively.}
	\label{fig2}
\end{figure}

Figures \ref{fig2}(a) and (b) show the enhancement maps of HMI continuum, CHASE \fe\ line center, CHASE continuum, and WST continuum emissions for two flaring times. Here the enhancement is defined as $(I-I_{0})/I_{0}$, where $I$ and $I_{0}$ are the intensity during flare time and the pre-flare intensity at a certain pixel, respectively. The uncertainties of these four wave-band enhancements are estimated to be about 1.6\%, 2.0\%, 1.3\%, and 2.4\%, respectively, by calculating the intensity fluctuations from a quiet-Sun region marked by the yellow dotted box in Figure \ref{fig1}(f). The average enhancements over the pixels above three times of the uncertainty as well as the maximum enhancements for these four emissions are given in Table \ref{tab:enhance}. The enhancement level of 20\% is also plotted in Figures \ref{fig2}(a) and (b) for each wave-band emission. From these maps we can see a few white-light brightening kernels with different enhancements, mainly located at the flare ribbons. We select three representative brightening kernels denoted by three colored plus symbols, among which two (green and magenta pluses) are located at different ribbons with opposite magnetic polarities in the south and the third one (blue plus) is near the big sunspot in the north. It seems that all three kernels are in the penumbra from different sunspots. At an earlier time ($\sim$17:50 UT, Figure \ref{fig2}(a)), the HMI continuum and CHASE \fe\ line emissions show relatively lower enhancements up to 62\% and 71\% with averages of 13\% and 21\%, respectively (listed in Table \ref{tab:enhance}). While the CHASE and WST continuum emissions exhibit much higher enhancements up to 184\% and 197\% with averages of 30\% and 56\%, respectively. Note that the maximum enhancements of HMI, CHASE, and WST continuum emissions at this time come from the same kernel as indicated by the magenta plus. At a later time ($\sim$17:52 UT, Figure \ref{fig2}(b)), the enhancements in all the emissions decrease notably (also see Table \ref{tab:enhance}), especially for the CHASE continuum. The maximum/average enhancements in the four wave bands are 43/11\%, 51/14\%, 75/15\%, and 135/36\%, respectively. Note that all the maximum enhancements are located at the same kernel denoted by the green plus. Generally speaking, the brightening kernels in the four wave bands are similar in morphology though having different enhancements.

\begin{table}[htb]
\caption{Maximum/Average Enhancements of the White-light Emissions in the X2.1 Flare}
\label{tab:enhance}
\centering
\begin{tabular}{ccccc}
\hline
\hline
Time & HMI continuum & CHASE \fe\ & CHASE continuum & WST continuum \\ 
 (UT)  & (\%) & (\%) & (\%) & (\%) \\
\tableline
 $\sim$17:50 & 62/13 & 71/21 & 184/30 & 197/56 \\
 $\sim$17:52 & 43/11 & 51/14 &   75/15 & 135/36 \\
\hline
\hline
\end{tabular}
\end{table}

Figures \ref{fig2}(c) and (d) plot the CHASE \fe\ and \ha\ line profiles from the three selected kernels at $\sim$17:50 UT as well as the temporal evolution of line profiles at the two most brightened kernels. From the second panel of Figure \ref{fig2}(c) we can see that the \fe\ profiles show a symmetric Gaussian shape with a shallower dip (compared with the background profile in gray) at the three kernels. In particular, the nearby continuum emission exhibits a significant increase at the two kernels marked by green and magenta pluses. Similarly, the \ha\ profiles show a prominently enhanced emission though with a central reversal at these kernels, as seen in the second panel of Figure \ref{fig2}(d). Note that the \ha\ line can show a blue or red asymmetry. For the temporal evolution, both the \fe\ and \ha\ lines display a notable increase followed by a decrease in the intensity at the two kernels marked by green and magenta pluses (see the third and forth panels in Figures \ref{fig2}(c) and (d)).

\subsection{The off-disk X1.5 Flare on 2023 August 7}

\subsubsection{Overview of the Flare Event}

\begin{figure}[htb]
	\centering
	\includegraphics[width=0.75\textwidth]{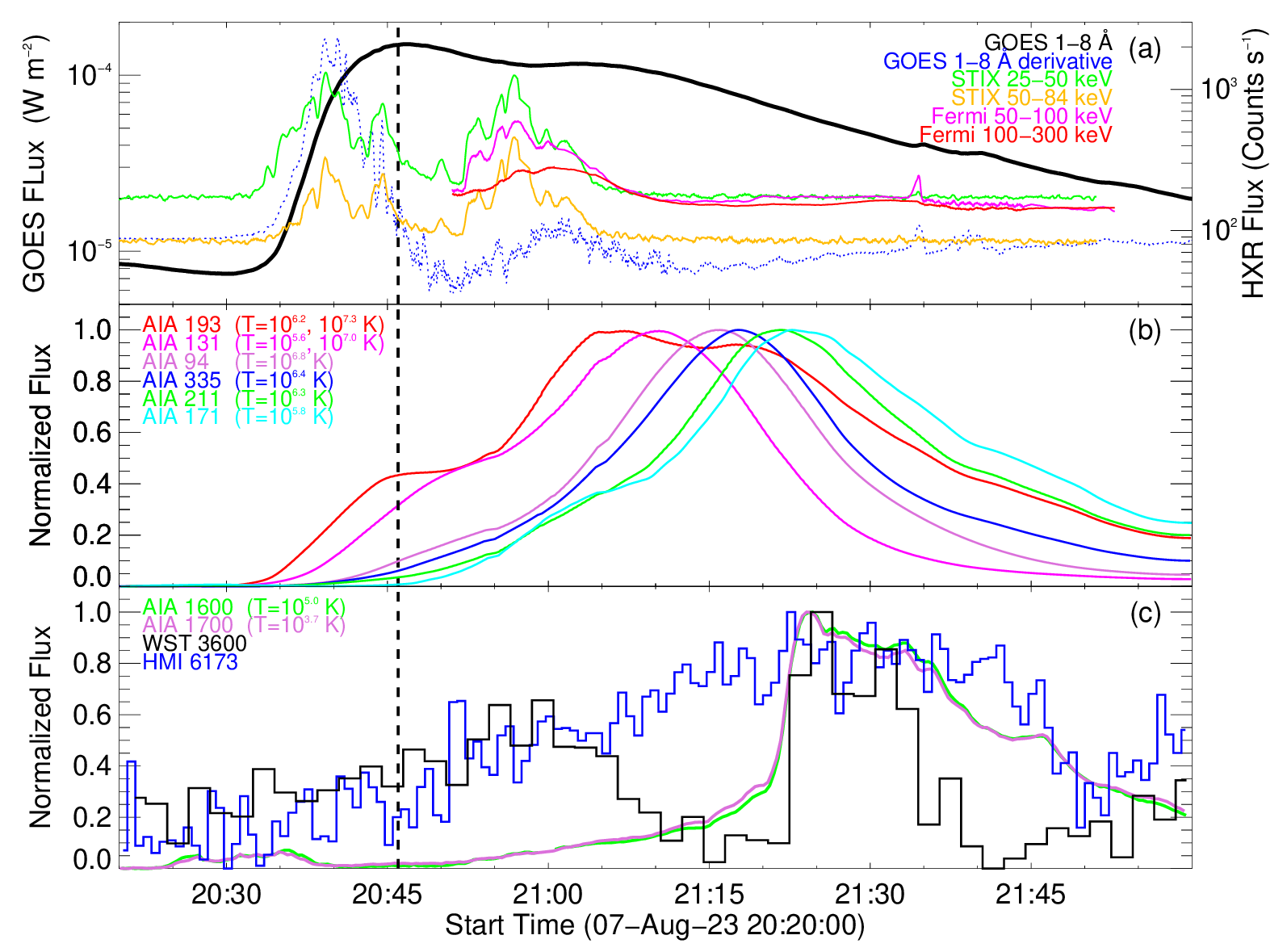}
        \includegraphics[width=0.75\textwidth]{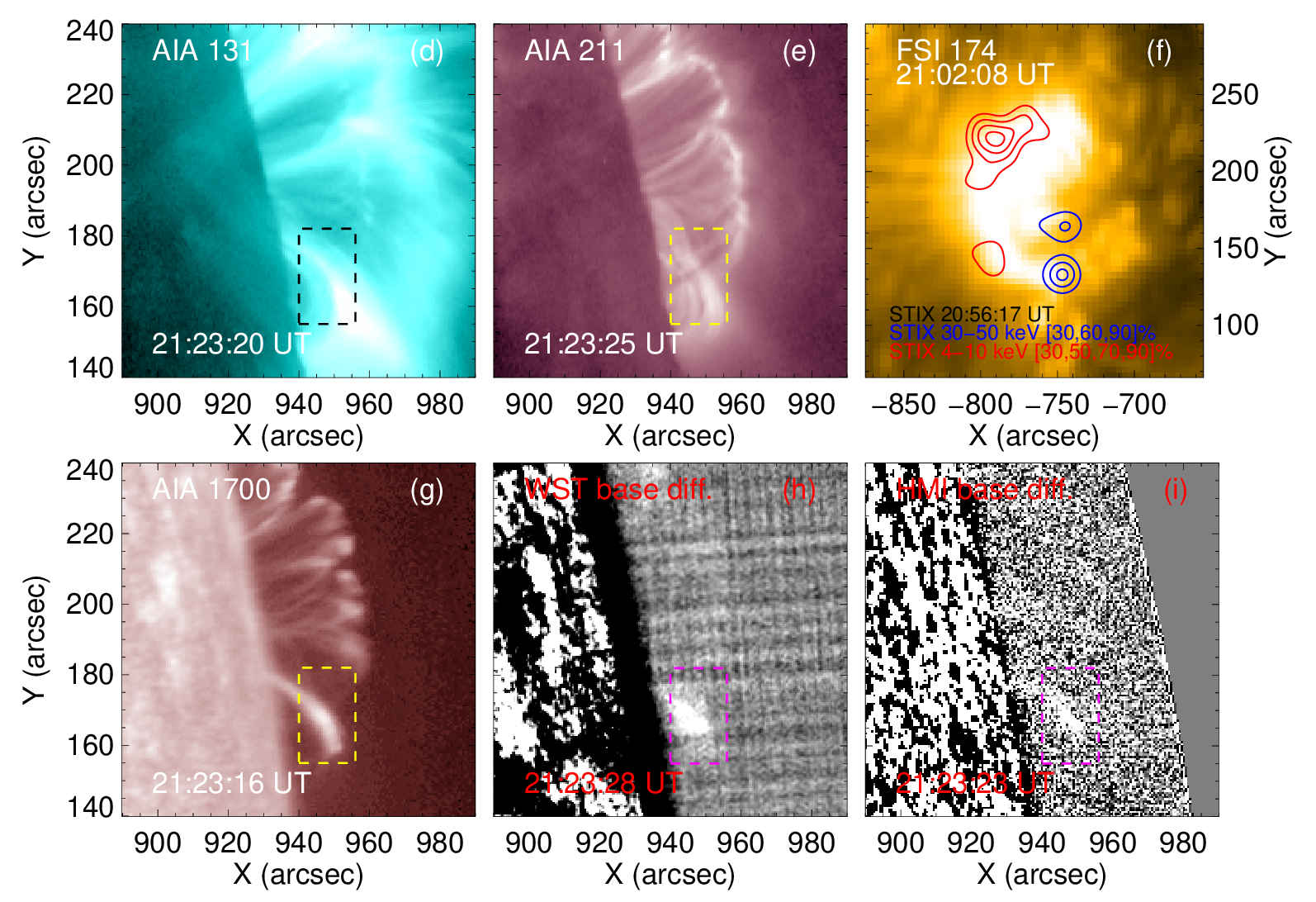}
	\caption{Overview of the off-disk X1.5 flare on 2023 August 7. (a) Light curves of GOES SXR 1--8 \AA, its time derivative, HXR fluxes from STIX and Fermi/GBM. (b) and (c) Multi-wavelength emission curves integrated over the loop region marked by the dashed box in panels (d), (e), and (g)--(i). The vertical dashed line indicates the flare peak time. (d)--(i) Multi-wavelength images for the flaring region during the decay phase. The AIA, WST, and HMI images are observed from Earth view while the EUI/FSI image at 174 \AA\ is obtained from another perspective. The red and blue contours in panel (f) indicate the loop and footpoint sources from STIX imaging. The dashed box in panels (d), (e), and (g)--(i) denotes the southern loops where show white-light emissions. The WST and HMI base-difference images are subtracted by a pre-flare image. The stripes visible above the limb in the WST image are caused by residuals of flat field.}
	\label{fig3}
\end{figure}

The X1.5 flare (SOL2023-08-07T20:46) occurred in NOAA AR 13386 (N12W88) at the west limb from Earth view, whose footpoints are occulted, with only flare loops visible above the limb (see Figures \ref{fig3}(d) and (e)). This off-disk flare started at 20:30 UT and peaked at 20:46 UT on 2023 August 7. Figure \ref{fig3}(a) shows the GOES SXR 1--8 \AA\ flux as well as HXR fluxes from Fermi and STIX. We can see that both SXR and HXR fluxes display two main peaks, indicative of two episodes of energy release. In addition, the STIX HXR fluxes match the SXR time derivative in the trend especially during the impulsive phase, i.e., the Neupert effect. It should be mentioned that STIX and EUI captured the flare loop as well as footpoint sources from another perspective as seen in Figure \ref{fig3}(f).

In this event, it is highlighted that the white-light emissions can evidently be seen on flare loops, particularly on the southern loops marked by the dashed box in Figures \ref{fig3}(d), (e), and (g)--(i). The multi-wavelength light curves integrated over the southern loops are plotted in Figures \ref{fig3}(b) and (c). One can clearly see a cooling trend from high- to low-temperature plasmas manifested in AIA light curves. The EUV emission peaks (from $\sim$21:05--21:22 UT) are followed by the peaks (at $\sim$21:25 UT) of UV emissions at AIA 1600 and 1700 \AA\ as well as the peaks (at $\sim$21:25 UT) of white-light continuum emissions from HMI and WST. A good match of peaks between the UV and white-light emissions indicates that the white-light emissions mainly originate from some relatively low-temperature plasmas.

\subsubsection{The White-light Emissions on the Flare Loops}

\begin{figure}[htb]
	\centering
	\includegraphics[width=1.0\textwidth]{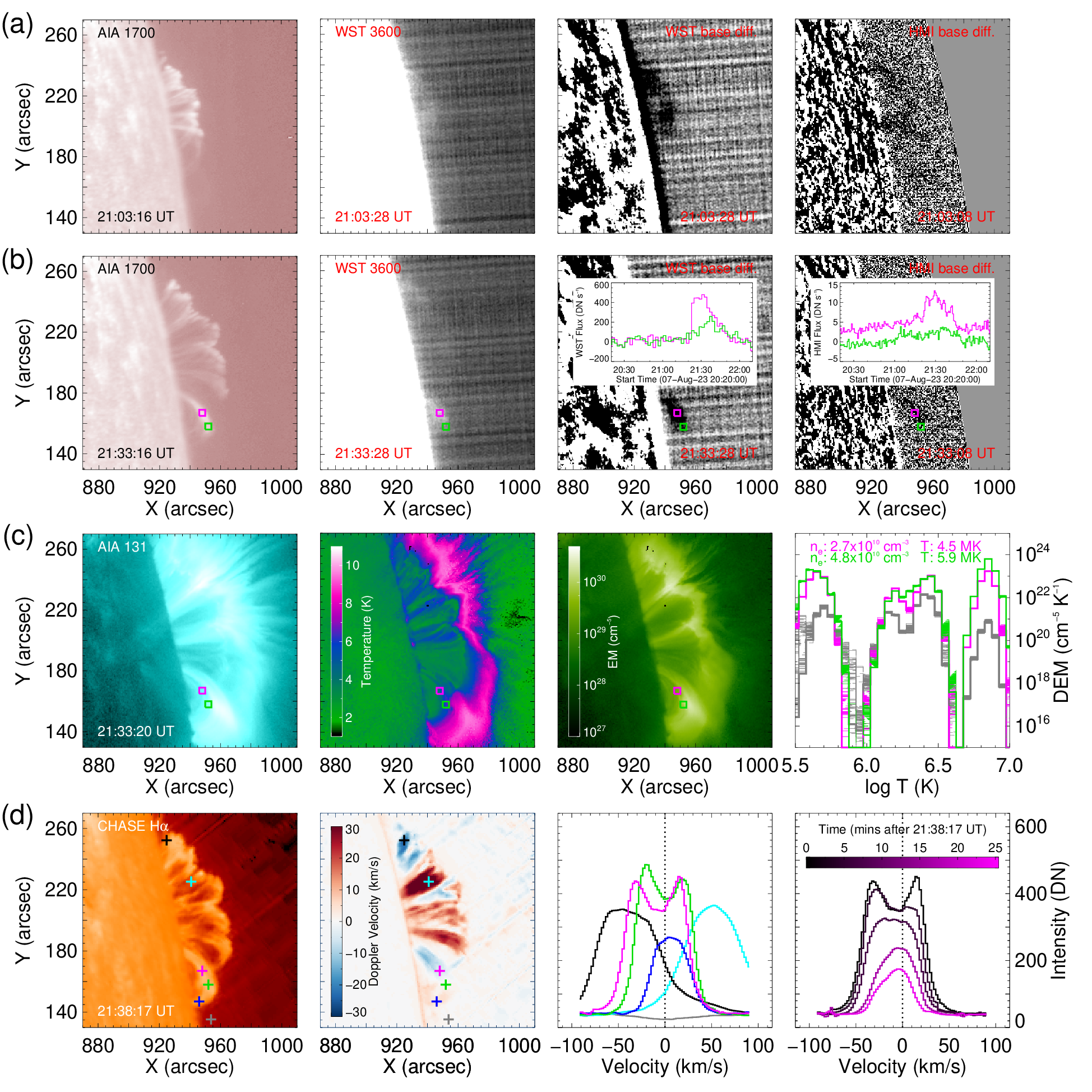}
	\caption{(a) and (b) AIA 1700 \AA, WST 3600 \AA, WST and HMI base-difference images at $\sim$21:03 and $\sim$21:33 UT for the X1.5 flare. The base-difference images are in a reversed color. The magenta and green squares in panel (b) indicate two selected loop locations showing white-light emissions, which are the same as the ones in panel (c). The insets in the third and fourth columns of panel (b) display the temporal evolution of the average intensities of WST and HMI continua at the two selected locations. (c) AIA 131 \AA\ image, temperature and EM maps at $\sim$21:33 UT, and DEM curves at the two loop locations marked by magenta and green squares. The gray DEM curve in the last column is from the location marked by green square before the flare for comparison. (d) CHASE \ha\ image, \ha\ Doppler velocity map, \ha\ line profiles at different locations marked by colored plus symbols in the image, and temporal evolution of \ha\ profiles at the location denoted by magenta plus.}
	\label{fig4}
\end{figure}

Figures \ref{fig4}(a) and (b) show the continuum images at AIA 1700 \AA, WST 3600 \AA, and HMI 6173 \AA\ during the decay phase of the flare. At an earlier time of $\sim$21:03 UT (Figure \ref{fig4}(a)), we can clearly see a cluster of flare loops with a height range of $\sim$19--28 Mm on the north in the AIA 1700 \AA\ image and some white-light emissions show up in WST and HMI images, corresponding to AIA loops. These white-light emissions are weak so as not so evident in the base-difference image of HMI. At a later time of $\sim$21:33 UT (Figure \ref{fig4}(b)), more flare loops appear in the AIA 1700 \AA\ image and the white-light emissions from WST and HMI can be seen mainly on the inclined southern loops that have a projection height of $\sim$13 Mm above the limb while a real height of $\sim$22 Mm after correction of the projection effect. We select two locations with each one having a size of 4\arcsec$\times$4\arcsec\ to display their average intensity evolutions for the HMI and WST continua (see the insets in the third and fourth panels of Figure \ref{fig4}(b)). The location marked by magenta square is on the loop part and the other denoted by green square is on top of the southern loops. We can see that the intensity curves from these two locations reach a maximum at $\sim$21:30 UT or after. It should be mentioned that the signal-to-noise ratios of HMI and WST emissions at the location denoted by magenta square are about 3 and 4, respectively.

We further make a differential emission measure (DEM) analysis for the flare loops at $\sim$21:33 UT using AIA EUV images \citep{Su2018,LiZT2022}, as shown in Figure \ref{fig4}(c). From the AIA 131 \AA\ image and temperature map, one can see some hot plasmas with an average temperature of $\sim$8 MK on the loop top. By contrast, the loops that show evident white-light emissions have a lower temperature and a lower emission measure (EM) or density (see the magenta and green squares in the maps). The DEM curves at the two locations are plotted in the last panel of Figure \ref{fig4}(c). It is seen that there contain high- and low-temperature components, with an average temperature of $\sim$5 MK. If assuming a line-of-sight length of $\sim$4 Mm (equivalent to the loop width), we can derive an electron density of $\sim$4$\times$10$^{10}$ cm$^{-3}$ at these loop locations.

Figure \ref{fig4}(d) displays the spectral features of flare loops at 21:38 UT from CHASE. Note that CHASE missed the main phase of the flare, starting the observation just from 21:38 UT, and that no evident emissions are found in CHASE \fe\ or the nearby continuum for this flare. From the \ha\ intensity map, one can see prominent cooling loops, especially the southern ones. We calculate the \ha\ Doppler velocity using a moment method. It is seen that some flare loops exhibit redshifts and some others show blueshifts, both of which are supposed to be caused by coronal rains, depending on the loop orientation. We also plot the \ha\ profiles on some loops showing different Doppler shifts. As seen in the third panel of Figure \ref{fig4}(d), the \ha\ line shows a redshifted emission profile on some loops and a blueshifted one on some others. Moreover, there are some \ha\ profiles showing a central reversal, say, the ones in magenta and green. These profiles are from the loops that show evident continuum emissions. In fact, the \ha\ profiles can change from a central reversal to a total emission over time (see the fourth panel of Figure \ref{fig4}(d)).


\section{Summary and Discussions}
\label{sec:summary}

In this Letter, we present on-disk as well as off-limb white-light emissions in two X-class flares observed by ASO-S/LST/WST, CHASE, and SDO/HMI. Among these white-light emissions, the Balmer continuum at 3600 \AA\ was rarely observed before, especially for the off-limb emissions from flare loops. Our results are summarized as follows.

\begin{enumerate}
\item The on-disk continuum emissions in the X2.1 flare are located at the flare ribbons, which peak at a similar time as the SXR time derivative and match the HXR emissions up to 300 keV in the time profile. At the brightening kernels, the WST 3600 \AA\ emission has a larger relative enhancement than the HMI 6173 \AA\ emission. The \fe\ line remains an absorption profile with the line center and nearby continuum enhanced prominently. The \ha\ line exhibits an emission profile with a central reversal and a blue or red asymmetry.
\item The off-limb continuum emissions in the X1.5 flare originate from the flare loops and mainly show up in the late decay phase. They peak around the same time with the AIA 1700 \AA\ emission, following a typical cooling trend of flare loops. The electron density on the loops showing white-light emissions is estimated to be $\sim$4$\times$ 10$^{10}$ cm$^{-3}$. On the loops, the \ha\ line displays an emission profile that can change from a central reversal to a total emission with time.
 \end{enumerate}

The white-light continuum emissions on flare loops during the X1.5 event follow a typical cooling trend from high- to low-temperature plasma. This indicates that the continuum emissions are associated with thermal plasma cooling. In addition, the electron density at the loops (in an order of $\sim$10$^{10}$ cm$^{-3}$) is found to be much lower than the density threshold of 10$^{12}$ cm$^{-3}$ \citep{Heinzel2017}, i.e., in the regime of Thomson scattering dominance. Thus, we consider that these continuum emissions are generated by Thomson scattering in principle, as already revealed in some previous studies \citep[e.g.,][]{Hiei1992}. This is worthwhile to be further investigated combining with a sophisticated radiative hydrodynamic simulation (Li et al., in preparation).

The white-light emissions at flare ribbons in the X2.1 event basically match the HXR emission at 100--300 keV in the time evolution, both of which follow the Neupert effect. This suggests that the white-light emissions are related to a nonthermal electron-beam heating either directly or indirectly. It is a pity that the HXR imaging for the footpoint sources cannot be obtained from Fermi/GBM or the Hard X-ray Imager (HXI; \citealt{Su2022}) on ASO-S, the latter of which was significantly influenced by some energetic particles from background at the observation time. We speculate that some of the white-light emissions can be directly caused by the electron precipitation, whereas the secondary effects, such as the radiative back-warming and chromospheric condensation, cannot be excluded in contributing to these ribbon emissions. According to the radiative hydrodynamic simulations in \cite{Allred2015}, the electron beams with an energy above 50--100 keV can deposit their energy in the middle to lower chromosphere, where the Balmer continuum around 3600 \AA\ is mainly formed during a flare (e.g., \citealt{Avrett1986,Fang1995}; Tian et al., in preparation). For the white-light emissions near two \fe\ lines from HMI and CHASE, they are mainly formed in the photosphere \citep[e.g.,][]{Hong2018,Hong2022}, and the radiative back-warming or some other mechanisms might be more important.

It is interesting to compare the on-disk X2.1 WLF under study with the X1.0 WLF studied by \cite{Song2023}. These two flares share similarities but have differences, too. Both flares have a similar relationship of the white-light emissions at ribbons with the HXR emission up to 300 keV, indicative of a nonthermal origin though as probably one of the heating sources. The \fe\ and \ha\ line profiles at white-light brightening kernels look similar in general for both flares. The former remains an absorption and the latter shows an emission with a central reversal during the flare time. The differences between the two flares are mainly from the values of white-light enhancements and the degree of asymmetry in the \ha\ line. The X2.1 flare in this study has much larger enhancements ($>$60\%) in the HMI and CHASE continua than the X1.0 flare ($\le$40\%) in \cite{Song2023}. For the line asymmetry, \ha\ shows a weak asymmetry in the X2.1 flare, whereas in the X1.0 flare it exhibits an obvious asymmetry. These differences are supposed to be primarily owing to different locations for the two flares. The X2.1 flare (N21W76) is close to the limb and the X1.0 flare (N17W49) is a little further away from the limb. Due to a limb-darkening effect, the region near the limb has a lower intensity that can cause a higher relative enhancement compared with the region further away. The weak and strong \ha\ asymmetries could be explained by different projection effects in these two flares.

\acknowledgments
We thank the anonymous referee very much for the constructive and valuable comments that helped to improve the manuscript. The ASO-S mission is supported by the Strategic Priority Research Program on Space Science, Chinese Academy of Sciences. The CHASE mission is supported by China National Space Administration. SDO is a mission of NASA's Living With a Star Program. Solar Orbiter is a space mission of international collaboration between ESA and NASA, operated by ESA. The STIX instrument is an international collaboration between Switzerland, Poland, France, Czech Republic, Germany, Austria, Ireland, and Italy. The EUI instrument was built by CSL, IAS, MPS, MSSL/UCL, PMOD/WRC, ROB, LCF/IO with funding from the Belgian Federal Science Policy Office; the Centre National d'Etudes Spatiales (CNES); the UK Space Agency (UKSA); the Bundesministerium f{\"u}r Wirtschaft und Energie (BMWi) through the Deutsches Zentrum f{\"u}r Luft- und Raumfahrt (DLR); and the Swiss Space Office (SSO). We thank Dr. Yang Su for helpful discussions. The authors are supported by NSFC under grants 12273115 and 12233012, the Strategic Priority Research Program of the Chinese Academy of Sciences under grant XDB0560000, the National Key R\&D Program of China under grant 2022YFF0503004, and NSFC under grants 12333009 and 11921003.

\bibliographystyle{apj}

\end{document}